\documentclass{epl}
\usepackage{graphicx}
\usepackage{bm}

\title{Experimental study of two-dimensional enstrophy cascade}
\author{G. Boffetta\inst{1} , A. Cenedese\inst{2}, 
        S. Espa\inst{2} \and S. Musacchio\inst{3}}
\institute{
  \inst{1} Dipartimento di Fisica Generale and INFN,
           Universit\`a degli Studi di Torino,
           Via Pietro Giuria 1, 10125, Torino, Italy \\
  \inst{2}  Dipartimento di Idraulica Trasporti e Strade, 
            Universit\`a La Sapienza,
            Via Eudossiana, 18 I-00184 Roma, Italy.\\
  \inst{3} Dipartimento di Fisica, Universit\`a La Sapienza, 
           P.le A. Moro 2, I-00185 Roma, Italy.
}
\pacs{74.27.Gs}{Isotropic turbulence; homogeneous turbulence}
\pacs{47.52.+j}{Chaos}

\begin{document}

\maketitle

\begin{abstract}
We study the direct enstrophy cascade 
in a two-dimensional flow generated
in an electromagnetically driven thin layer of fluid.
Due to the presence of bottom friction,
the energy spectrum deviates 
from the classical Kraichnan prediction $k^{-3}$. 
We find that the correction to the spectral slope depends 
on the thickness on the layer, in agreement with a theoretical
prediction based on the analogy with passive scalar statistics.
\end{abstract}

\section{Introduction}
Laboratory studies of two-dimensional turbulence are affected by
the interaction with the three-dimensional environment in which they
are embedded. This interaction is often frictional and generates
an additional linear damping term to the equation of motion.
A well known example, which is the object of the present letter,
is the flow in a shallow layer of fluid, 
which is subject to the friction with the 
bottom of the tank where it is contained \cite{PT97}. 
Another example is the motion of a soap film dumped from the
interaction with the surrounding air \cite{RW00}.
Linear friction dumping is not restricted to laboratory experiments:
an important example is the Ekman friction in geophysical 
fluid dynamics \cite{salmon}.

As predicted in a remarkable paper by Kraichnan in 1967 \cite{K67}, 
when a two-dimensional layer of fluids is forced at intermediate
scales, smaller than the domain-size and larger than the viscous 
dissipative length-scales, two different inertial ranges are observed.  
This is a consequence of the coupled conservation of
energy (mean square velocity) and enstrophy (mean square vorticity)
in the inviscid limit, a basic difference with respect to 
three-dimensional turbulence phenomenology.
The energy flows toward the large scales giving rise to an
inverse energy cascade characterized by the $k^{-5/3}$ 
Kolmogorov spectrum. 
Small scale statistics is governed by a direct enstrophy cascade
which is expected to develop a smooth flow with $k^{-3}$ energy spectrum,
with a possible logarithmic correction \cite{K71}.

The presence of a linear friction term affects the direct cascade
in a dramatic way: the enstrophy flux across scales is no longer
constant and the scaling exponent of the 
spectrum differs from the Kraichnan prediction $k^{-3}$ by
a correction proportional to the friction 
intensity \cite{B00,NOAG00}.
The origin of this correction can be understood exploiting  
the analogy between two-dimensional vorticity field and a scalar field
passively advected by a smooth velocity field \cite{BCMV02}.
For the case of a passive scalar with a finite lifetime  
it is possible to obtain explicit expressions 
for the scaling exponents of the power spectrum and structure functions
which depend on the intensity of the linear dumping and on the 
statistics of finite-time Lyapunov exponents \cite{C98,NAGO99}.

In this letter we experimentally study the statistic of 
the direct enstrophy cascade in a two-dimensional flow generated
in an electromagnetically driven thin layer of fluid.
This experimental setup has been successfully used for studying
several problems connected to two-dimensional turbulence
\cite{PT97,S86,CHZ03}. 
Here we show that the friction with the bottom of the tank 
induces a correction on the spectral slope at large wave-numbers, 
in good agreement with theoretical predictions \cite{B00,NOAG00}. 


\section{Experiment}
The experimental apparatus consists of a squared Plexiglas tank 
whose dimensions are $50 \times 50$ cm. 

The tank is filled with two layers of fluid, 
subsequentially injected from the bottom. 
The upper layer of fresh water is placed over 
a bottom layer of an electrolyte 
solution of water and NaCl (density $\sim 1060 g/l $).   
At the end of the filling procedure we obtain a stable stratification 
with controlled fluids thickness. In all the experiments 
the fluid thickness $h_1$ for the lower fluid has been maintained equal to 
$0.3 cm$ while the thickness of the upper fluid $h_2$ varies between 
$0.3$ and $0.7 cm$.
The total thickness of the two layers will be denoted by $h$.

\begin{figure}
\centering
\includegraphics[draft=false, scale=0.37, clip=true]{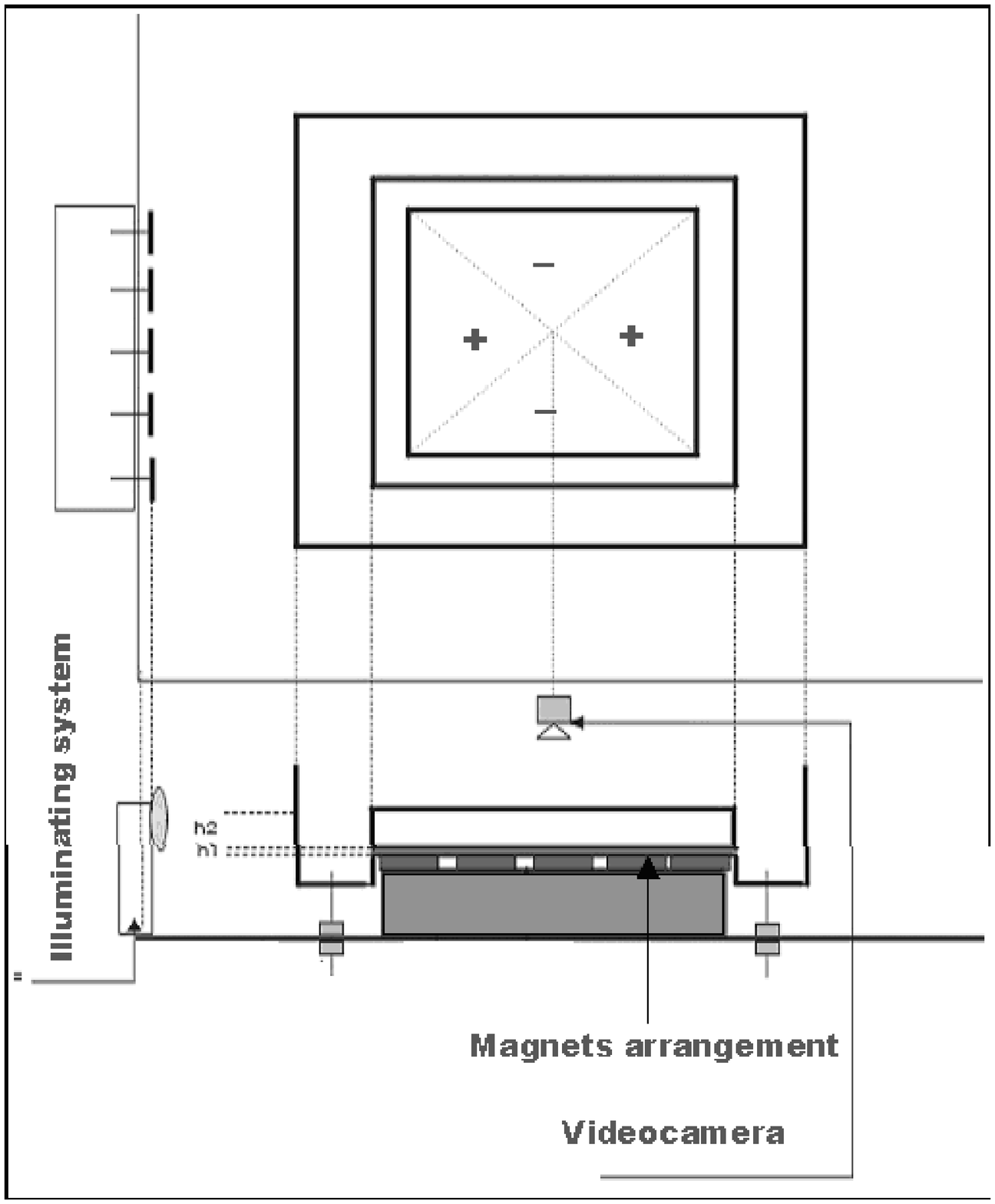}
\includegraphics[draft=false, width=7 truecm, 
height=7 truecm, clip=true]{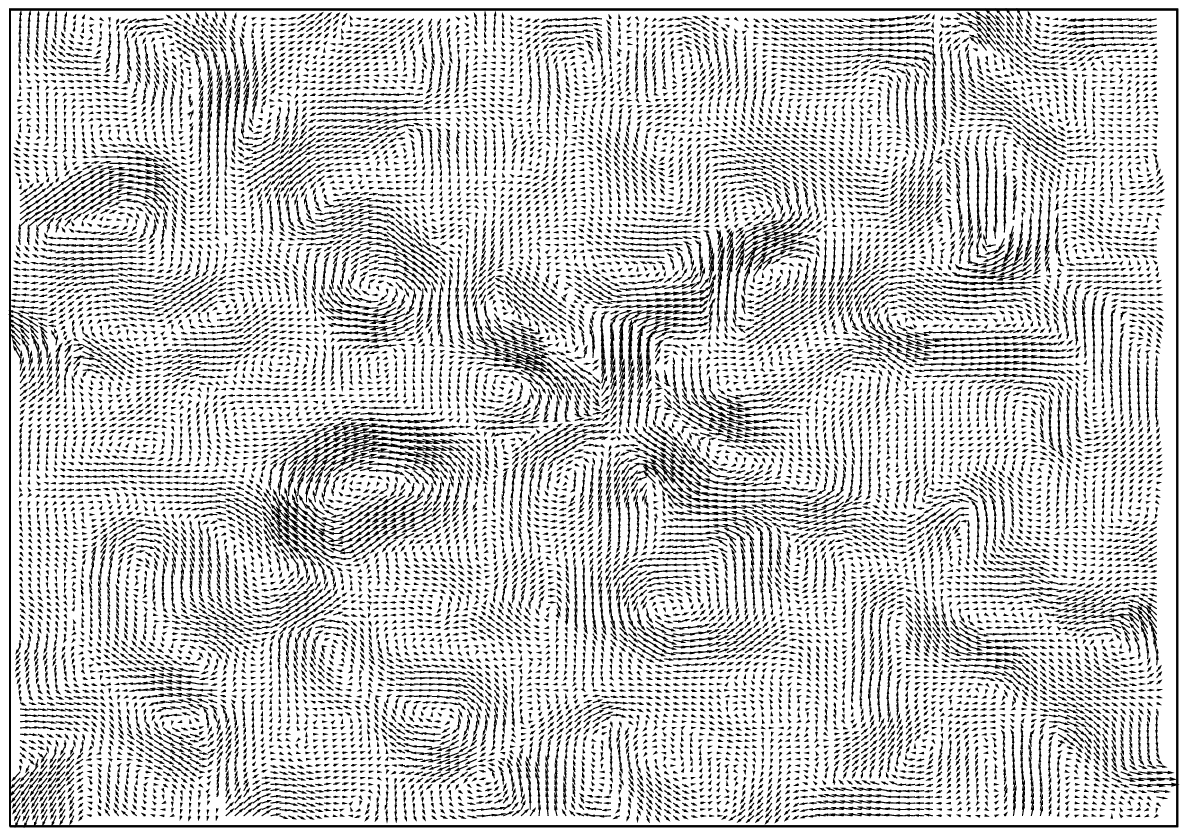}
\caption{
a) A schematic upper view and side view of the experimental setup.
b) Snapshot of the velocity field reconstructed by the FT
technique on the regular at resolution $128 \times 128$ 
grid for the run $3$
(see Table~\ref{table1}).
}
\label{fig:1}
\end{figure}

Neodymium permanent magnets are placed below the bottom of the tank
and disposed in four triangles with magnetic field in 
the vertical direction and alternating sign (see Fig.\ref{fig:1}).
An electric current, horizontally driven through the cell, interacts
with the magnetic field and moves the NaCl solution via the Lorenz force. 
The forcing current is generated by a computer controlled power supply
which provides voltage signal of fixed amplitude 
and randomly alternating direction. 
The correlation time for voltage sign reversal is tipically $4s$.  
The combined action of electric and magnetic forcing on the NaCl solution
induces the continuous formation of opposite signed vortical structures 
whose characteristic length-scale $L_{f}$ is related to the distance 
between opposite-signed magnets 
and whose characteristic time-scale is of the order of voltage sign reversal. 

The free fluid surface is seeded using tiny buoyant styrene particles 
(with typical size $d \sim 250 \mu m$), 
and the test section is illuminated using 
an array of lamps placed orthogonally to one side of the tank. 
The fluid flow is recorded using a standard speed video camera.
A maximum duration of 6 minutes is chosen for each experiment, to ensure 
that both density stratification and two-dimensionality are maintained 
\cite{CMT94}. 
The velocity field has been reconstructed by image analysis based on 
a Feature Tracking (FT) approach\cite{EC05}. This tracking procedure allows for 
higher seeding densities than classical Particle Tracking Velocimetry, 
and provides an
accurate reconstruction of a large number (almost 20000) of 
Lagrangian trajectories. The interpolated Eulerian velocity field is 
therefore highly detailed, maximizing the information content of raw data.
In Fig.~\ref{fig:1} we show an example of instantaneous velocity 
field reconstructed at resolution $128 \times 128$. 
\begin{figure}[t]
\centering
\includegraphics[draft=false, scale=0.7, clip=true]{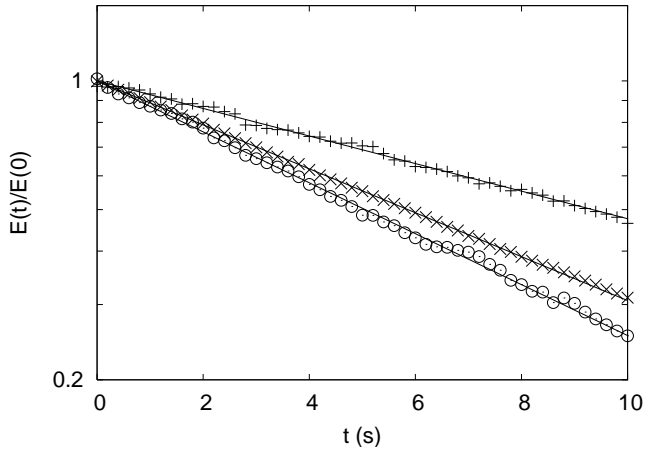}
\caption{ 
Decay of kinetic energy for the unforced runs
(run $1$ $(+)$, run $2$ $(\times)$, run $3$ $(\circ)$). 
The lines represent exponential fits which give the 
friction coefficients reported in Table~\ref{table1}. 
}
\label{fig:3}
\end{figure}
\begin{table}[b]
\caption{Parameters of the experiments. 
Friction coefficient $\alpha$, root-mean-square velocity $u_{rms}$, 
root-mean-square vorticity $\omega_{rms}$, spectral index 
correction $\xi$}
\label{table1}
\begin{center}
\begin{tabular}{cccccccccccccccc}
Run \# & $h (cm)$ & $\alpha (s^{-1})$ & $u_{rms} (cm s^{-1})$ & $\omega_{rms} (s^{-1})$ & $\xi$\\
\hline 
1 & 0.8 & 0.0371 & 1.324 & 0.747  & 0.49 \\
2 & 0.9 & 0.0591 & 1.330 & 0.640  & 0.78 \\
3 & 1.0 & 0.0686 & 0.790 & 0.602  & 1.02 \\
\hline 
\end{tabular}
\end{center}
\end{table}
\section{Theory and experimental results}

The dynamics of a thin layer of fluid electromagnetically forced 
is described by the linear dumped 2D Navier-Stokes equations, 
which can be written for the vorticity $\omega={\bm \nabla} \times {\bm v}$ as
\begin{equation}
{\partial \omega \over \partial t} + {\bm v} \cdot {\bm \nabla} \omega=
\nu \nabla^2 \omega - \alpha \omega + f(t)
\label{eq:1}
\end{equation}
where $\nu$ is the fluid viscosity, $f(t)$ is the time-dependent 
external forcing.
The bottom drag is parameterized by the linear friction term 
$-\alpha \omega$. Assuming a Poiseuille-like 
vertical velocity profile, the intensity of 
the friction coefficient can be related to the 
total thickness of the layer $h$ as    
$\alpha \propto \nu/h^2$ \cite{SCVCH01}. 
\begin{figure}[t]
\centering
\includegraphics[draft=false, scale=0.7, clip=true]{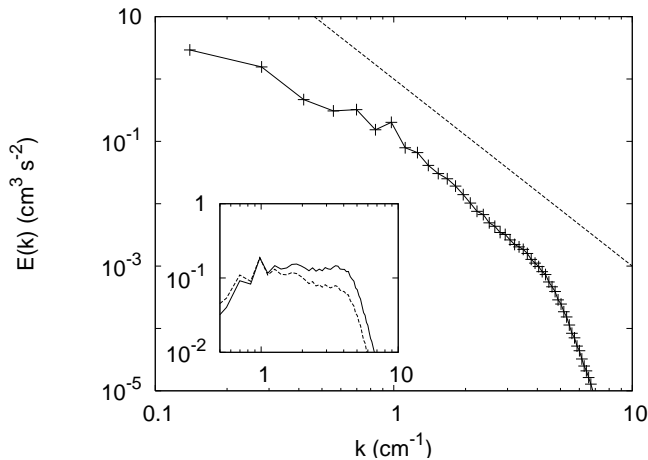}
\caption{ 
Kinetic energy spectrum for the run $1$ ($+$),   
The dashed line represents the Kraichnan prediction $k^{-3}$.
In the inset we show the spectrum compensated with $k^{-3-\xi}$
(solid line) with exponent $\xi$ given in Table \ref{table1}
and with the Kraichnan prediction $k^{-3}$ (dashed line).
}
\label{fig:5}
\end{figure}

In the inviscid-unforced limit ($\alpha=\nu=f=0$), the NS equations
(\ref{eq:1}) conserves both the kinetic energy $E=1/2 \langle v^2 \rangle$
and the enstrophy $Z=1/2 \langle \omega^2 \rangle$ \cite{KM80}.
For $\alpha>0$ the enstrophy cascading to small scales is removed
by friction, allowing to disregard the viscous term in (\ref{eq:1})\cite{B00}.
In this limit, both the energy and the enstrophy decay exponentially in the
unforced case with a decaying characteristic time $\tau = 1/(2 \alpha)$
\begin{equation}
E(t) = E(0) e^{- 2 \alpha t} \;\;\;\; Z(t) = Z(0) e^{- 2 \alpha t}
\label{eq:2}
\end{equation}
In Figure \ref{fig:3} we plot the decay of the total energy
for three experiments with different total thickness $h$ 
starting from time at which the electric forcing is switched off.
The agreement with the exponential decay is remarkable and allow 
a direct measurement of the friction coefficient $\alpha$. 
Further this is a confirmation {\it a posteriori} 
of the irrelevance of the viscous term.

The remarkable prediction made in \cite{NOAG00} is that 
for any $\alpha>0$ the Kraichnan scaling exponent in the direct 
cascade has a correction proportional to $\alpha$. The 
argument is based on an analogy with the dynamics of a
scalar with finite lifetime $\tau=\alpha^{-1}$ passively 
transported by a smooth flow, which is governed by an equation 
formally identically to (\ref{eq:1}) \cite{NLHT00}.
We remark that the extension of the passive scalar argument
\cite{NAGO99} to the {\em active} vorticity field is not trivial,
as $\omega$ determines the velocity field. For completenes, in the
following we report the ``mean field'' derivation of the exponent
correction, a complete derivation for the active case can be
found in \cite{BCMV02}.
\begin{figure}[t]
\centering
\includegraphics[draft=false, scale=0.7, clip=true]{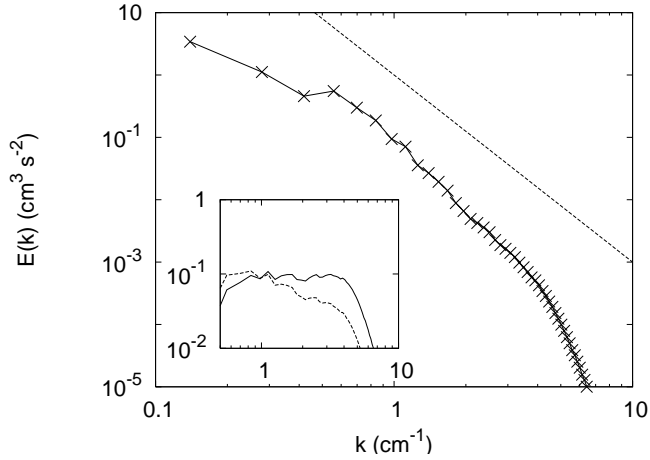}
\caption{ 
Kinetic energy spectrum for the run $2$ ($\times$).
The dashed line represents the Kraichnan prediction $k^{-3}$.
In the inset we show the spectrum compensated with $k^{-3-\xi}$
(solid line) with exponent $\xi$ given in Table \ref{table1}
and with the Kraichnan prediction $k^{-3}$ (dashed line).
}
\label{fig:6}
\end{figure}

Let us consider a fluctuation of vorticity 
$\delta \omega(L_f,0)=\Omega$ generated by the 
forcing at scale $L_{f}$ and time $0$, {\it i.e.} a blob of vorticity 
of size $L_{f}$ at the rms value $\Omega$.
Due to the chaotic flow, the blob is exponentially 
stretched with a mean rate given by the Lyapunov exponent $\lambda$. 
Because of the incompressibility of the velocity field, 
after a time $t$ the blob is contracted in the transverse dimension 
to a scale $r = L_{f} \exp(-\lambda t)$.  
This is the mechanism for the {\it direct cascade} phenomenon, 
{\it i.e} the fluctuation has been transported  
from the large injection scale $L_{f}$ to the small scale $r$.
Taking into account the decay induced by the friction we can write
\begin{equation}
\delta \omega(r,t) \sim \delta \omega(L_f,0) e^{-\alpha t} \sim
\Omega \left({r \over L_f} \right)^{\alpha/\lambda}
\label{eq:3}
\end{equation}
In the statistically steady regime sustained by the forcing, 
this argument provides the scaling exponent for 
the second-order vorticity structure function: 
\begin{equation}
S_{2} (r) \equiv \langle (\delta \omega (r))^2 \rangle \sim 
\Omega^2 \left({r \over L_f} \right)^{2 \alpha/\lambda}
\label{eq:3.1}
\end{equation}
and thus the scaling exponent of the enstrophy spectrum 
$Z(k) \sim k^{-1-2 \alpha/\lambda}$.
Finally one obtain the mean field prediction for the energy spectrum 
\cite{NOAG00}
\begin{equation}
E(k) \sim k^{-3-\xi}
\label{eq:4}
\end{equation}
with correction exponent $\xi=2 \alpha/\lambda$.
We remark that for $\alpha>0$ the correction gives an energy
spectrum steeper that $k^{-3}$ which is a posteriori consistent with the
assumption of smooth velocity field.

A more refined version of this ``mean-field'' argument, which takes
into account the fluctuations of the Lyapunov exponents \cite{BJPV98},
can be made \cite{NOAG00,BCMV02}. The result is again a correction
which makes the spectrum steeper than the Kraichnan prediction.
An extensive numerical study of two-dimensional Navier-Stokes equations
with friction has confirmed the above theoretical picture \cite{BCMV02}.

\begin{figure}[t]
\centering
\includegraphics[draft=false, scale=0.7, clip=true]{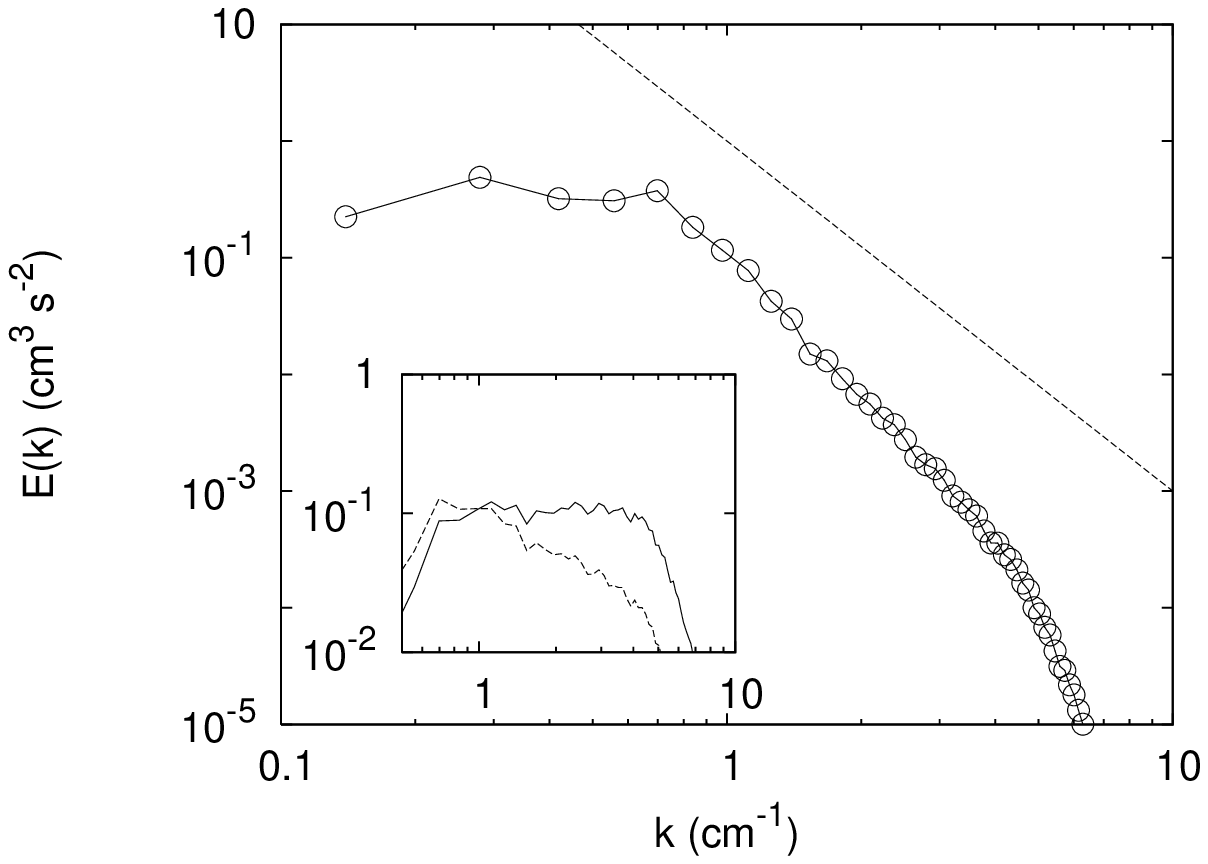}
\caption{ 
Kinetic energy spectrum for the run $3$ ($\circ$).
The dashed line represents the Kraichnan prediction $k^{-3}$.
In the inset we show the spectrum compensated with $k^{-3-\xi}$
(solid line) with exponent $\xi$ given in Table \ref{table1}
and with the Kraichnan prediction $k^{-3}$ (dashed line).
}
\label{fig:7}
\end{figure}

Figures \ref{fig:5}, \ref{fig:6} and \ref{fig:7} 
shows the energy spectra obtained from 
Fourier transform of the FT velocity data for three
experiments with different total thickness in stationary conditions. 
The spectra are obtained by averaging over about
$50$ realizations of the velocity field.
A clear cascade range with power-law scaling is evident 
in intermediate wave-numbers $0.7 cm^{-1} < k < 4.0 cm^{-1}$. 
The forcing wavenumber $k_{f} \simeq 0.7 cm^{-1}$ 
corresponds to an injection scale $L_{l} \simeq 9\,cm$, 
consistent with the size of the array of magnets. 

A fit of the energy spectra in the range $1.0 cm^{-1} < k < 4.0 cm^{-1}$
gives the exponent corrections $\xi\simeq 0.49$, 
$\xi \simeq 0.78$ and $\xi \simeq 1.02$
for the three runs respectively. A direct comparison with the
theoretical prediction (\ref{eq:4}) would require the knowledge 
of the Lyapunov exponent of the flow. 
It is possible to give a simple estimation 
by considering the characteristic time in the direct cascade which is
given by $\omega_{rms}^{-1}$. Therefore, we can assume that 
the Lyapunov exponent of the flow is proportional to $\omega_{rms}$,
a quantity which is easily determined from the velocity field.
Finally, by considering a couple of different runs we have:
\begin{equation}
\frac{\xi_2}{\xi_1} = \frac{\alpha_2}{\alpha_1} \frac{\lambda_1}{\lambda_2}
= \frac{\alpha_2}{\alpha_1} \frac{\omega_1}{\omega_2}
\label{eq:5}
\end{equation}
Using the values of $\alpha_i$ and $\omega_i$ from table \ref{table1}
we obtain the predictions 
$\xi_1 / \xi_3  =0.81$ and $\xi_2 / \xi_3  = 0.44$
which are close to the direct estimation from the spectra 
$\xi_1 / \xi_3 = 0.78$ and $\xi_1 / \xi_3 = 0.49$. 

\section{Conclusion}

In this letter we have presented an experimental study of
the effects of bottom friction on the direct enstrophy cascade 
observed in thin layer of fluid electromagnetically forced.
Direct measurements of the friction coefficient 
are obtained from the exponential decay of the total 
energy when the forcing is switched off. 
In the stationary forced case the energy spectra 
of the reconstructed velocity fields display a power law behavior, 
with a slope $k^{-3-\xi}$ which differs from the
Kraichnan prediction $k^{-3}$.   
The correction $\xi$ to the spectral slope 
is due to the friction exerted by the bottom wall on the fluid    
and increases with the friction intensity.
Its value can be predicted by theoretical arguments, in good agreement 
with our experimental measurements. 
Being the correction roughly proportional to the square of the inverse 
of the total thickness of the fluid, the observed effect
is expected to be extremely relevant in the case of thin layers.

\acknowledgments
This work has been supported by Italian MIUR.


\end{document}